\RequirePackage[OT1]{fontenc}

\documentclass[10pt, conference, compsocconf]{class/IEEEtran}

\usepackage{iftex}

\newcommand{\fixheaderbadness}{
	
	\hbadness=10000 
	\vbadness=10000 
	\hfuzz=3000pt 
	\pretolerance=10000
	\tolerance=10000
	
	\usepackage[safe]{silence} 
	
	\WarningFilter{latexfont}{undefined}
	\WarningFilter{latexfont}{Font shape}
	\WarningFilter{latexfont}{Some font shape}
	\WarningFilter{latex}{Unused global option(s)}
	\WarningFilter*{latexfont}{Font shape}
	\WarningFilter*{latexfont}{Font}
	
}

\newcommand{\insertimage}[4][1.0]{

	\begin{figure}[ht]
		\centering
		\includegraphics[width=#1\textwidth]{#2}
		\caption{#3}
		\label{#4}
	\end{figure}
	\FloatBarrier
	
}

\global\def\typewriterstyle{false}

\newcommand{\settypewriter}[1] {
	
	\ignorespaces\lowercase{\global\def\typewriterstyle{#1}}\unskip
}

\global\def\trueOption{true}

\global\def\falseOption{false}

\newcommand{\setdocumentimports}{

	\ifx\typewriterstyle\trueOption 
				
		\ifLuaTeX
			\usepackage{typewriter}
		\fi
	\fi
}{}

\CLASSINFOpdffalse

\newcommand{\imageformat}{
	
	\ifCLASSINFOpdf
	
		\usepackage[pdftex]{graphicx}
		\usepackage{epstopdf}
		
		\DeclareGraphicsExtensions{.pdf,.jpeg,.png}
	
	\else
	
		\usepackage[dvips]{graphicx}   
		\DeclareGraphicsExtensions{.eps}
	
	\fi
}

\newcommand{\importelvishifexists}{
	
	\ifPDFTeX
	
		\IfFileExists{tolkienfonts.sty}{
			
			\ifx\typewriterstyle\falseOption
			\pdfmapfile{=tolkien.map}
			\fi
			
			\usepackage{tolkienfonts}
			
			\global\def\elvish{true}
			
		}{
			
			\global\def\elvish{false}
		}
	
	\fi

}

\newcommand{\elvishlanguage}[1]{

	\ifx\elvish\trueOption
	
		\begin{quenya}
			#1
		\end{quenya}

	\fi
}

\newcommand{\elvishfootnote}[1]{
	\relax
	\ifx\elvish\trueOption
		\relax\footnote{\elvishlanguage{#1}}
	\fi
}

\ifPDFTeX
	\settypewriter{false}
\fi

\ifLuaTeX
	\settypewriter{true}
\fi

\importelvishifexists

\imageformat

\usepackage[cmex10]{amsmath} 
\usepackage{eso-pic}

\setdocumentimports

\usepackage{algorithmic}
\usepackage{array}
\usepackage{mdwmath}
\usepackage{mdwtab}
\usepackage{eqparbox}
\usepackage[caption=false,font=footnotesize]{subfig}
\usepackage{cite}
\usepackage{pdftexcmds}
\usepackage{catchfile}
\usepackage{ifluatex}
\usepackage{placeins}

\usepackage{todonotes}
\usepackage{url}

\fixheaderbadness

\begin{document}

\title{Blockchain and radio communications over suborbital spaceflights: Watchtowers and Mystics}

\author{\IEEEauthorblockN{
    Sergio Martinez-Losa del Rincon\IEEEauthorrefmark{1},
    Esteban Damian Gutierrez Mlot\IEEEauthorrefmark{2}
    \IEEEauthorblockA{\IEEEauthorrefmark{1}Email:  sergiomtzlosa@protonmail.com}
    \IEEEauthorblockA{\IEEEauthorrefmark{2}Email:  esguti@protonmail.com}
  }
}


\maketitle

\begin{abstract}
  The communication between ground and satellites is tough due to long distances and interferences in the propagation medium. Using the blockchain technology and a series of ground nodes, information can be distributed in a secure and reliable way. The satellite discovers nearby ground stations and sends the information that it is verified with other nodes present in the same network. Ground nodes are classified into two types, master (Watchtowers) and working (Mystics) nodes. The Watchtower controls the entering gate of new nodes to the network and the distribution of information between the working nodes.
  
The Mystics nodes use a verification process with the information. Each Mystic exchanges the information with the n-nearest nodes and receives tokens as a reward is the information has been verified using a proof-of-work algorithm.
\end{abstract}


%
\IEEEpeerreviewmaketitle

\section{Introduction}
Blockchain has been a disruptive technology in different fields. It is well known, the use of Blockchain in the cryptocurrencies~\cite{Nakamoto2008}, but there are other fields, like robotics~\cite{Mushtaq2019}, healthcare~\cite{Abdullah2019}, software license validation~\cite{Bayhan2018} and more~\cite{AbuNaser2019}. In our research process, we have not found any application in the field of aeronautics, in particular, suborbital devices, field whereby this paper presents the use of Blockchain and their advantages. The main advantages of Blockchain are decentralization and the immutability of the data. These two advantages perfectly fit with long flight communications like communications between a weather satellite and ground stations. 

The information in this kind of communications is transmitted primarily over the air and all over extensive areas. An extensive area of coverage needs on the ground a wide number of receptors collecting the information transmitted by the device. Blockchain creates a decentralized network of trusted nodes on the ground. The nodes can belong to different companies or even countries and thanks to blockchain, they can cooperate and trust each other, because the ledger of information shared cannot be stolen, corrupted or influenced by anyone. 

In this paper, it is presented a novel method of Blockchain application in the fields of High Altitude Balloon (HAB) communications, that could be extended to other suborbital spaceflights communication. The method consists of maintaining a Blockchain Database (BDB) made of diverse nodes on land. The nodes synchronize the BDB when the HAB passes by their sphere of action. The BDB is filled with sensor data collected by the HAB. The consensus mechanism proposed for the participating ground nodes is the well-known Proof of Work (PoW). Although PoW has been used in our demonstration, we leave open the possibility of using other consensus mechanisms such as the novelty Proof of Stack (PoS)~\cite{conf/sscc/BarhanpureBD18} or the much-studied Practical Byzantine Fault Tolerance (PBFT)~\cite{OSDI99*173}.

The source code of the demonstration is made available in a public repository~\cite{blockchain-satellite}.

\section{Blockchain and low-orbit satellites}
In the dawn of the machines, there was no communication over then, later on, a set of protocols and different devices arise to interact between machines and humans.

The blockchain technology can be applied to low-orbit devices to send the data in a feasibly and reasonably way. This technology ensures that data integrity is performed and also it can create a gratification token to users to maintain the network alive and make it grow.

The protocol described in this paper is based on two kinds of devices: the main device, hereinafter, the ``Watchtower'' and working nodes called ``Mystics''.

\subsection{Networking: Watchtower and Mystics}

Every suborbital device belongs to one ``Circle''. A Circle is the collection of devices and networks which receive, distribute and process the information related to one suborbital device.

Inside the Circle, the Watchtower is the gatekeeper. It has an active role in the network allowing or denying a node to be part of the Circle.

Not only nodes are controlled by the Watchtower, but also information can be shared between Circles through Watchtowers, making possible to communicate different suborbital devices in an isolated way.

Due to the high number of nodes that may become part of a Circle, it is allowed to have more than one Watchtower in the same network, avoiding bottlenecks.

A Mystic is in charge of processing and storing the information. Every Mystic must be registered on the Watchtower before entering the Circle, which means Mystics must supply credentials to a Watchtower.

All Mystics are connected between them and with the Watchtower. A Mystic can belong only to one Circle but can switch to another Circle leaving the previous one.

\insertimage[0.4]{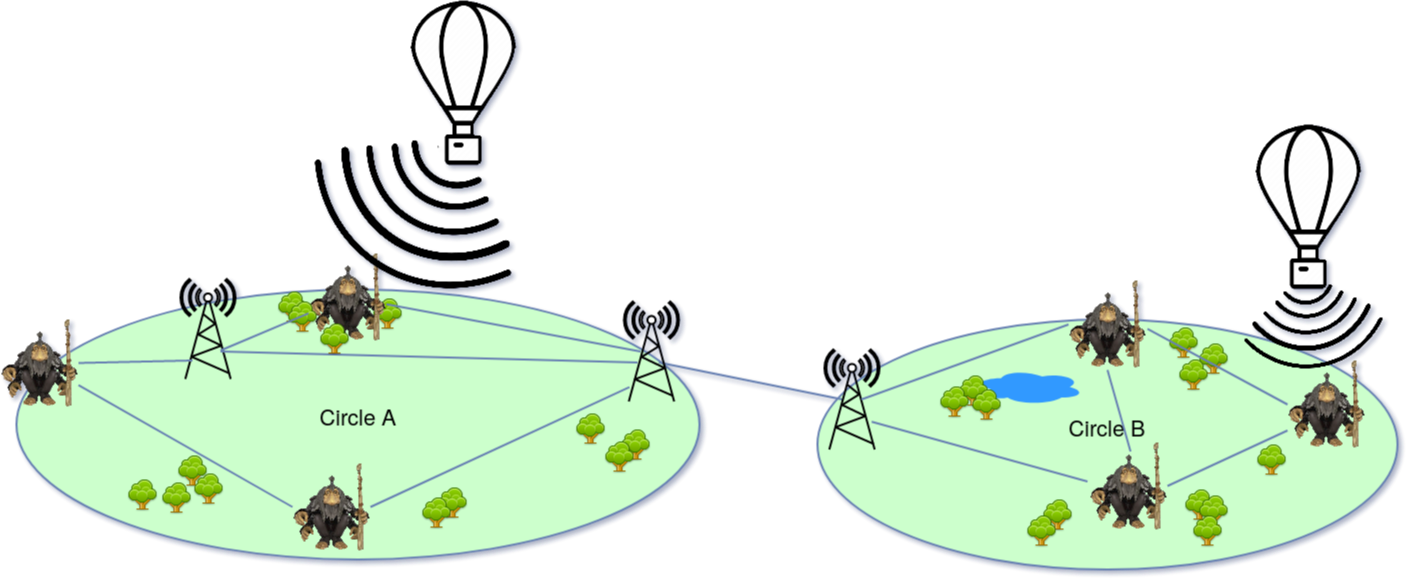}{Mystics, Watchtowers and Circles.}{fi:circles}

\subsection{Witchcraft and sorcery over data payload}

This section will explain the process to obtain and distribute to all over the Circle from a particular Mystics.

Let's think that a suborbital device~\cite{Andrusenko2005, Dinh2015} (for instance: low-orbit balloon) collecting data from several sensors and sending them through a known (HTTPS) or unknow protocol to our Circle.

Each time the low-orbit satellite wants to send new data, it sends a broadcast message to find Mystics and later on, from the returned response the satellite decides with one is the nearest from its current position. Once the low-orbit device selects the Mystic, it sends the data. Then, the data will be inside the Circle.

Each Mystic inserts the received data and gets the hash resulting in the insertion in its blockchain. The Mystics send the calculated hash to the closest Watchtower. The Watchtower will order the hashes and select the most recent one. This hash will be sent back to the Mystics which gave hashes to the Watchtower.

Each Mystic updates its own stored data by talking and exchanging information with the n-nearest Mystics in a data verification process, thus all Mystics have the same information. On this verification process, each Mystic receives a gratification token called ``Sparkling''. There is a delay time of 15 minutes between each Mystic-to-Mystic update. The data will be considered as verified when 80\% of the circle or more verify the data.

As additional security operation, after data verification, the Watchtower asks to n-random number of Mystics for their last hash, if the hashes between the Watchtower and the Mystics match, the Mystics have been updated accurately.

This process can be repeated on each circle, it is important to know that the satellites can only discover one or more circles if the satellite has the Circle identification, this identification must be a long, unique and hexadecimal token generated on the ground before landing.

There are more ways to identify a remote device against a Circle, the proposed way is one of them but there are many others like name and password, two-factor authentication, asymmetric cryptography~\cite{journals/compsec/Laun92}, or LDAP server. The method proposed in this paper has been selected because is the simplest way to authenticate consuming the lowest power possible.

The Circle token cannot be shared and it is generated per satellite and flight. The identification must be on board when the low-orbit device leaves the ground because it will not be possible to update it during the flight. 

\subsection{Data update between Mystics}

All Mystics are connected in the same circle. A Mystic sends the last data available on its blockchain to the n-nearest Mystics. After the insertion of data, the Mystics match the token. If the tokens are equal, then the data is verified. This process is repeated on each Mystic until reaching the 80\% (or more) of the available Mystics on the circle.

If the data cannot be verified by the Mystics, the data insertion is revoked and information discarded.

\subsection{Communication between Watchtowers of different circles}

The Watchtowers are the gateway between circles. They exchange trust tokens that allow external Mystics to login into another circle, these tokens can be revoked at any time.

The token is built with an exchange of credentials between the Watchtowers. Let's explain the token exchange with an example:

\begin{enumerate}
	
	\item Watchtower A (WA) wants to enter into circle B (CirB) lead by Watchtower B (WB).
	\item WA sends a request to WB and WB responds with a token revocable after 15 min.
	\item if WA wants more than 15 minutes, it will perform a second request using the current temporal token and requesting a time window for n-minutes.
	\item If the 15 minutes token expires the WA can request another 15 tokens to WB.
	
\end{enumerate}

\subsection{Black Mystics: Sorcery forks}

It is not intended to have a fork on the circle as the information only enter using only one Mystic. It is a future task to know how behaves the circle over sorcery forks.

\subsection{Mystics lifecycle: crossing the life abyss}

The Watchtower sends discovering packets (charm) every n-minutes to know if all registered Mystics are alive. 

Once the lifecycle of a Mystic is over and it crosses the life abyss, the Mystic stops making sorcery at the circle and the Watchtower removes the Mystics from the alive Mystics list\elvishfootnote{Everybody believes I am crazy because meanwhile they live in the past I am given them a future} to the abyss list. A Mystic can back again to the Circle performing again a registering request. The Watchtower has the information of the dead-and-gone Mystics and can return them to life in the circle.

\subsection{Circle security enhancement}

It is not implemented yet a method to avoid the 51\% attack~\cite{Natoli2016, Keenan2017, Bae2018} over the network. This circumstance sets the limit of external Mystics from a foreign circle in (n/2)-2 where n-item is the number of Mystics from our circle. This limit adds up 2 foreign Mystics to our circle, 2 instead of 1 to avert fails. It is a future task to find a better system to avoid this kind of attack.

\section{Demo for a Proof of Concept}

This demo shows the proposed system as a proof of concept. Although the system needs dozens of Mystics on the ground and at  Watchtower to show the true potential of this idea, we are going to show a low-scale infrastructure just for demonstration purposes.

The demonstration consists of a Circle with 3 Mystics and a Watchtower. The Circle is isolated from other circles and obtains the data from an external source (for instance, a weather balloon which is a low-orbit device).

We have uploaded the source code to a public repository~\cite{blockchain-satellite}. Have in mind that the code is in continuous development and could be incomplete, but currently, the code is stable and there is also a docker setup to test the full blockchain system, check the repository out for further details.

\subsection{Hardware proposed}

The hardware list for this proof of concept is the following:

\begin{itemize}
	\item 3 Raspberry Pi 3B+ to work as Mystics
	\item 1 Raspberry Pi 3B+ to work as Watchtower
	\item Ethernet cable Cat6
	\item Router with internet access
	\item A low-orbit device with sensors
\end{itemize}

\subsection{Software proposed}

The low-level system to store the data is a non-relational database because this kind of database has good performance with blockchain technology~\cite{Karafiloski2017, Lemieux2017}. In particular, in this demo, we will use MongoDB~\cite{mongodb} as it easy to use and it fits our needs.

As the communication protocol between the external source (satellite) and the Circle, we choose HTTPS, because it is a lightweight and well-know communications protocol.

It is also necessary for a set of Web Services to send the information to the circle, then it is used PHP as a programming language to build these services. The service will be onto an Apache2 server

The payload to send is not defined by now but we will use JSON~\cite{journals/computer/Severance12c} notation to create a payload structure.

The security related to external nodes is build-in with a time-token system. This token is used by external Mystics which want to enter the circle. There is also a firewall to avoid DoS attacks~\cite{Asare2019, Alkurdi2018}, just necessary to open a port for the Web Services.

\subsection{Demo explanation and results}

The objective of this demo is to know how behaves a homemade blockchain network with time-tokens security.

In this circle, all Mystics are connected between them and each Mystic is connected to the Watchtower.

This proof of concept covers a fully working blockchain in Rust language with user security and data encryption over the blockchain.

The application is optimized for low-capacity devices and performs in a good way on the latest computers.

This blockchain uses a sha256 encryption of \textit{/etc/passwd} salted with a fixed hash (random hash) plus another hash chunk from the md5 \textit{/etc/passwd} file, that means when the user changes the password, the data cannot be recovered as the original \textit{/etc/passwd} file is modified, the security of the data is bound to a physical file.

The hashes from each block of the blockchain are verified using the Merkle root hash using the hashes from six ancestors blocks if the block is verified the integrity of data is lost.

The blockchain is stored into a MongoDB database, this technology gives a big amount of records to store on a low-orbit journey.

The demo performed well accurately and the results were what we expected. The only issue found, was that the time verification of the data inserted in the blockchain is quite high (around 5 seconds). This issue is due to the low-powered hardware used in the demo.

The data were secure and the Watchtower performs well within this infrastructure.

\section{Conclusion and future work}

The system has to be tested with a bigger circle with more Mystics and Watchtowers interacting between them. A bigger circle will let us achieve more accurate results.

The demo proposed is low-scaled, and maybe the results will be different in a larger network. The non-relational database gives little delay with few nodes on the network, it is a future work to find a better solution or try with other non-relational databases with better performance.

The WebServices performed well and we cannot measure high delays on communications over low-orbit devices, in the future we will study a replacement to Web Services as an alternative communication technology.

In conclusion, the communications between satellites and ground stations can be faster and near real-time with blockchain technology, it is in our hands to develop a new way to obtain a big amount of data from the current system. A bigger set of data will set us into a new edge of possibilities and challenges regarding the keep and distribution of the data.

\bibliographystyle{class/IEEEbib}
\bibliography{blockchain-hab}

\end{document}